\documentstyle [12pt]  {article}

 \evensidemargin\oddsidemargin

\begin{document}
 \title{{\bf FUZZY SPACE-TIME, QUANTIZATION   \\
 AND GAUGE INVARIANCE \\ }}
\author {{\bf S.N.Mayburov} \\
% \thanks{E-mail ~~ maybur@sgi.lpi.msk.su  ~~
 Lebedev Inst. of Physics\\
   Leninsky Prospect 53,  Moscow, Russia, RU-117924\\
  e-mail: mayburov@sci.lebedev.ru \\}
 \date { }
 \maketitle
%
%\begin{abstract}

% \maketitle

\begin{abstract}
 Dodson-Zeeman fuzzy topology considered as the possible
  mathematical framework of  quantum geometric formalism.
 In such formalism the states of massive
 particle $m$ correspond to  elements of fuzzy manifold
 called  fuzzy points. Due to their weak (partial) ordering, $m$
space coordinate $x$ acquires principal uncertainty $\sigma_x$.
It's shown that $m$ evolution
 on such manifold corresponds to quantum dynamics.
  It's argued also that particle's interactions on such fuzzy manifold
should be gauge invariant.
\end{abstract}

% \keywords{Fuzzy Topology
% \and Quantization}

\section{Introduction}
\label{intro}

There are several serious reasons to try to formulate quantum
mechanics (QM) in geometric terms. For instance, general
relativity is essentially geometric theory, but the attempts to
quantize gravity suffer the serious difficulty already at
axiomatics level. Such formalism  can be useful also for
development of gauge field theory, which is also mainly geometric;
its implications can be important for the analysis of QM
foundations.
% Importance of geometric methods in quantum physics is duly
%acknowledged now, however the reformulation of quantum mechanics
%(QM) in geometric terms is still an open problem ( see \cite
% {Mar} and refs. therein). Such formalism can be applied primarily
% in quantum gravity realm but also for gauge field theory; its
% implications can be important
% for the analysis of QM foundations.
 Currently, the main impact of QM geometrization studies is done
  on the exploit of Hilbert manifolds (\cite {Mar} and refs. therein), however, the
  results obtained up to now have quite abstract form, and their
  applicability  to particular problems is questionable.
Alternatively, in approach considered here the basic structure is
the real manifold equipped with fuzzy topology (FT) \cite
{Dod,Zee,Sos}.
% The popular example of it is Connes
%noncommutative geometry which attempts to describe the
%fundamental interactions at Planck distances \cite {Con}.
 In connection with such mathematical framework it's
worth to mention  the
  noncommutative fuzzy spaces with
 finite (sphere, tori) and infinite discrete structure
 \cite {Bal}.
The general feature of such theories  is that the space
coordinates turn out to be principally fuzzy, the reason of that
is the noncommutativity of coordinate observables $x_{1,2,3}$.

Meanwhile, the  similar coordinate fuzziness  exists for the
manifolds equipped with dedicated FT
 \cite {Dod,Zee}.
% Basing on its principles,
%  the consistent novel geometry with fuzzy features was
% formulated.
% hence it's instructive to study  what kind of
% physical theory such topologies induces \cite {May2,Vax}.
Earlier, it was argued that in its framework the quantization
procedure  itself can be defined
 as the transition from the classical  phase space to fuzzy one \cite {May2,Vax}.
 Therefore, in such approach
the quantum properties of particles and fields are deduced
directly from
 the geometry of  phase space induced by underlying FT and don't need to be  postulated
 separately of it.
% \cite {May2,Vax}.
% In particular, the
% space coordinate uncertainty is the consequence of  fuzzy
% properties of such space geometry.
 In particular, in such formalism the system evolution can be
 described as the geometrodynamics which is equivalent to
 quantum dynamics \cite {May2,Vax}; as the  example, the dynamics of massive
particles will be  considered.
% Besides the standard QM noncommutativity of $x, p_x$ observables,
% such theories postulate that the orthogonal space coordinates
% $x,y,z$  also don't commute. In such terms we study the
% commutative fuzzy geometry, because in our formalism the
% noncommutativity of any pair of phase space coordinates isn't
% assumed beforehand,it can appear (or not) only as the result of
% derivations.
% Here we shall investigate mainly the novel
% symmetries of fuzzy manifold and related physical states. The
% resulting constraints on the interactions between particles are
%obtained;
Previously,  some phenomenological assumptions were used by the
author for its derivation, here the  simple formalism which
permits to avoid them is described, its main features can be
found in \cite {May3}.
%   Note that the fuzzy structures were
% used earlier for the development of QM axiomatic in operator
% algebra setting, yet in such formalism the quantum dynamics is
% always postulated, no attempts to derive it were published \cite
% {Pyk}.
It was argued also that the particle interactions on such fuzzy
manifold possess the local gauge invariance \cite {Vax}.
% and under simple
%assumptions would correspond to Yang-Mills theory \cite {Vax}.

   Note that the fuzzy structures were
used earlier for the development of QM axiomatic in operator
algebra setting  \cite {Pyk}, yet in such formalism the quantum
dynamics is always postulated, no attempts to derive it were
published. The important example, illustrating the connection
between fuzzy structures and quantum dynamics described in \cite
{Ali}.
% As was shaon earlier, the interactions on such fuzzy
%manifold can possess the local gauge invariance and under simple
% assumptions would correspond to Yang-Mills theory \cite {Vax}.

 \section {Topological Fuzzy Structures}

Here the main FT features important for the construction of
dynamics
   on fuzzy manifold  are reviewed \cite {Dod,Sos}. For the
start we consider the  geometry  for which its fundamental set is
unambiguously defined, later this assumption, in fact, will be
dropped. To illustrate FT formalism
 let's consider it first for some discrete
space.
 If its fundamental set  $S$ is  totally ordered set,
 then for its elements
 $\{a_i\}$  the  ordering relation between the
element pairs $a_k\leq a_l$ (or vice versa) is fulfilled.
% For example the element $a_k\in P_A$ ordered relative to some $a_j$
% but isn't ordered relative to some others
But if $S$ is the  partially ordered set (Poset), then
 some its element pairs can
 enjoy the  incomparability  relations (IR) between
 them: \(a_j \sim a_k\). If this is the case, then both $a_j \le
a_k$ and $a_k \le a_j$ propositions
 are false, and such  structure acquires some nontrivial properties
\cite {Schr}. For instance, consider 2-dimensional discrete plane
$D$ with elements $d_{ij}=\{x_i,y_j\}$ where all $x_i,y_i$ are
integer numbers. Suppose that the ordering relation is defined
from the comparison of both $d$ coordinates, i.e. $d_{ij} \le
d_{kl}$ iff $x_i \le x_k. \, and \, .y_j \le y_l$. Then if such
relation isn't fulfilled or for $x$ coordinate or for $y$ (but
not for both of them simultaneously),
 it means that $d_{ij} \sim d_{kl}$       \cite {Schr}.

 As further  example, important for our formalism, consider
 poset $S=A^p \cup B$, which includes the
 subset of 'incomparable' elements $A^p=\{ a_j \} $,
and ordered subset $B=\{b_i \}$.
% which is maximal $S$ subset for
%all element pairs of which the ordering relations hold.
 For the
simplicity we suppose that in $B$ the element's indexes grow
correspondingly to their ordering, so that $\forall \, i$,  $b_i
\le b_{i+1}$.
% Any $b_j$ is incomparable at least to one $a_i$.
Let's  consider  $B$ interval  $\{ b_{l},b_n\}$
% such that $b_j \in BS_{ln}$ if $b_l \le b_j \le b_n$,
%%  , i.e. $D^T$ subset for which
%%  $\forall a_i, b_j; \, a_l \le a_i, b_j \le a_{l+n}; \, n \ge 2$.
 and suppose that   some $A^p$ element $a_j$ is confined in
  $\{b_{l},b_n\}$, i.e.
 %  $a_j \in BS_{ln}$, i.e.
 %   $\forall k \ge 1 \,$:
 $b_{l} \le a_j \, ; \, a_j \le b_{n}$,
 and  simultaneously  $a_j$ is incomparable with all internal $\{b_{l}, b_n\} $
 elements: $b_i \sim a_j;\,$
   $\forall \, i \,; \, l+1\le i \le n-1 $. In this
case $a_j$  can be regarded as 'smeared' over such interval,
which is the rough analogue of $a_j$ coordinate uncertainty
relative to $B$ 'coordinate', if to consider the sequence of $B$
elements as the analogue of  coordinate axe. The generalization
of poset structure is the tolerance space for which the ordering
relations can be nontransitive, the similar property possesses
some quantum structures \cite {Zee,Dub}.

It's possible to detalize the described smearing introducing the
fuzzy relations, for that purpose one can put in correspondence
to each $a_j, b_i$ pair of $S$ set the weight $w^j_i \ge 0$
 with the norm $\sum_i w^j_i=1$.
% for $a_j,a_i$ or $b_j, b_i$ pairs it' unimportant.
%% If $b_0$ is ordered in$A$, for example $b_0=a_i$, then $w^0_j=\delta_{ij}$
%% If $w$ is defined for all $a_i,b_j$ pairs, their total
In this case $S$ is  fuzzy  set, $A^p$ elements $\{a_j\}$ called
the fuzzy points (FP) \cite {Zee,Gott}. For the example considered
above, one can ascribe $w^j_i=\frac{1}{n-l+1}$ to all $b_i$ inside
$\{b_{l}, b_n\} $ interval,  $w^j_i=0$ for other $b_i$.
% The simple example is the homogeneous incomparability:
%  $w^j_i=\frac{1}{n}$ for
% $a_i \in [a_l,a_{l+n}]$;
% $w^j_i=0$ outside of it.
%% It can be interpreted as $b_j$ homogeneous smearing
%%  inside $[a_l,a_{l+n}]$.
%% as the discrete analogue of space coordinate $x$ uncertainty.
%% Note that in distinction from
%% the regarded case, in general an arbitrary Foset isn't necessarily Poset.
%%
In its simplest form the continuous fuzzy set $C^F$ is defined
analogously to discrete one: $C^F=A^P \cup X$ where $A^P$ is the
same discrete subset,
% the ordered subset $A$ is substituted
$X$ is the continuous ordered subset, which is
 equivalent to $R^1$  real number axe.
  Correspondingly, fuzzy relations between elements $a_j,x$ are
described by real function
 $w^j(x)\ge 0$ with the norm $\int w^j dx=1$.
The ordered point $x_a$ is characterized in this framework by
$w^a(x)=\delta(x-x_a)$.
%% in this case, $C^F$ is called the fuzzy space.
%%  If on $X$ some metrics is defined, $C^F$ is called fuzzy space, below we
%% shall regard $C^F$ only as the sum of 1-dimensional Euclidean space and
%% the discrete set $B$ of the fuzzy points.
Remind that in 1-dimensional Euclidean geometry,
 the elements of its manifold $X$ are the points
$x_a$ which constitute the ordered continuum set.
% $x_a$ position
% on $X$  in 1-dimensional Euclidian geometry is characterized by
%the real number $x_a$ in some coordinate system.
 Yet in
1-dimensional geometry equipped with FT the position of fuzzy
point $a_j$ becomes the positive normalized function $w^j(x)$ on
$X$; $w^j$ dispersion $\sigma_x$ characterizes $a_j$ coordinate
uncertainty on $X$. If metric is defined on $X$ then $C^f$ is
called fuzzy manifold.
 Note that FT realm incorporates several
alternative formalisms in which different FP definitions are
exploited, we use here the one given in \cite {Gott}, in fact,
it's the geometric analogue of real fuzzy number \cite {Dub}.

 We shall suppose that the  geometry of physical world corresponds
 to geometry equipped with FT considered here. Note that
 in such  formalism $w^j(x)$ doesn't have any probabilistic
 meaning but only the algebraic one,
 characterizing the properties of fuzzy value $\tilde {x}_j$.
 To describe the distinction between the fuzzy structure and
probabilistic one, the correlation $\kappa_0 (x,x')$ defined over
$w_j$ support can be introduced; thus if $w_j(x_{1,2}) \ne 0$,
then $\forall x_{1},x_{2}$; $\kappa_0 (x_{1},x_{2})=1$ for FP
$a_j$ and $\kappa_0 (x_{1},x_{2})=0$ for probabilistic $a_j$
distribution. Thus $a_j$ state $G$ on $X$ is described by two
functions $G=\{w_j(x),\kappa_0 (x,x')\}$.
%which characterize the  fuzzy value $\tilde {x}_j$.
As will be shown below, the similar bilocal correlations describe
 the dynamical properties of physical objects.

%\section {Linear Model of Fuzzy Dynamics}

\section{Linear Model of Fuzzy Dynamics  }
\label{sec-1}

In the described framework  the massive particle of 1-dimensional
classical mechanics corresponds to the ordered point $ {x_a}(t)
\in X$. By the analogy, we suppose that in 1-dimensional fuzzy
mechanics (FM) the particle $m$
 corresponds to  fuzzy point $a(t)$ in $C^F$
% ( its modification for 3-dimensions will be regarded in final chapter).
 characterized by normalized positive density $w(x,t)$.
% Here we'll
% attempt to calculate $m$ free evolution from the geometric
% premises only.
Beside $w(x)$,
$m$  fuzzy state $|g\}$ can also depend on
  other $m$ degrees of freedom (DFs)  characterizing its evolution.
%  The obvious one is
% $\frac{\partial w}{\partial t}$, yet it's more convenient to
% replace it by some other DF,
To illustrate it, consider $m$ average velocity:
\begin {equation}
\bar{v}= \frac{\partial}{\partial
t}\int\limits^{\infty}_{-\infty}xw(x)dx=
\int\limits^{\infty}_{-\infty}x \frac{\partial w}{\partial
t}(x,t)dx                                    \label {AB}
\end {equation}
It's reasonable to assume that $\bar v(t)$ is independent of
$w(x,t)$, below we shall  look for such additional DFs in form of
real functions  $q_{1,...,n}(x,t)$.
%  and  $w$ flow  velocity  $v(x,t)$.
% If $w \in C^2$, then by definition:
% \begin {equation}
%                v(x)= -\frac{1}{w(x)}\int\limits^{x}_{-\infty} \frac{\partial %w}{\partial t}(\xi) d\xi        \label {AZ}
% \end {equation}
% Plainly, $\bar {v}(t)$ is average $m$ velocity.
Let's  suppose that in FM $m$  evolution
% because it
% should be invariant relative to space and time shifts
 is local, i.e.:
 \begin {equation}
                  \frac{\partial w}{\partial t}(x,t)=-\Phi(w, q_1,...,q_n)     \label {AXY}
 \end {equation}
where $\Phi$ is an arbitrary  function which  depends on $w(x,t)$,
$q_{1,...,n}(x,t)$ only.
% Such locality is
% essential assumption of our theory, however, the alternative
% nonlocal  variants are quite complicated and demand to use the
% additional constants.
From $w$ norm conservation it follows that:
 \begin {equation}
   \int\limits^{\infty}_{-\infty} \Phi(x,t)dx = -\int\limits^{\infty}_{-\infty} \frac{\partial w}{\partial t}(x,t)dx=
       -\frac{\partial }{\partial t}\int\limits^{\infty}_{-\infty}
       w(x,t)dx = 0
                         \label {AZZZ}
 \end {equation}
%Since  $w$ free evolution should possess $x, t$-shift invariance,
% $\Phi$ can't depend on $x,t$
% directly, but only on $w(x,t)$ and  $q_i(x,t)$.
% yet it's natural  property for fuzzy object $m$.
 If to substitute: $\Phi=\frac{\partial J}{\partial x}$
 where $J(x)$ is some differentiable function, then eq. (\ref {AZZZ})
 demands:
 \begin {equation}
 J(\infty,t)-J(-\infty,t)=0    \label {X}
 \end {equation}
%$ From $X$ reflection invariance it's sensible to settle:
%We suppose that this relation is fulfilled
 If  such equality is
 fulfilled, then $J(x)$ can be regarded as $w$
 flow, and eq. (\ref {AXY}) is equivalent to
  $1$-dimensional flow continuity  equation \cite {Lan}:
\begin {equation}
                  \frac{\partial w}{\partial t}=- \frac{\partial J}{\partial x}     \label {AZ1}
 \end {equation}
% Below it will be
% shown that in our theory\ $J(\pm \infty,t)=0$.
Meanwhile, $J(x)$ can be decomposed formally as:
  $J=w(x)v(x)$ where $v(x)$  corresponds to $1$-dimensional $w$ flow
velocity \cite {Lan}.
% $J_0$ is nonclassical term which is supposedly depends on $w,v$ and
%obeys to relation:
% $\int\limits^{\infty}_{-\infty}J_0dx=0$.
 In these terms eq.(\ref{AZ1})
can be written as:
 \begin {equation}
                  \frac{\partial w}{\partial t}= -v  \frac{\partial w}{\partial x}
                   -  \frac{\partial v}{\partial x} w   \label {AZ2}
 \end {equation}
We shall assume that $v(x)$ can be considered as novel $m$ DF.
% Thus in  1-dimensional FM $w$ evolution obeys  to $1$-dimensional
% flow continuity  equation, which derived here just from the
% locality axiom.
% The analogous equation in classical kinetics or
% hydrodynamics are obtained  from additional assumptions about the
% properties of evolving media \cite {Lan}.
  Note that for normalized density $w(x,t)$ the condition expressed by
   eq.  (\ref{X}) is
  trivial, in particular, it's fulfilled if $w$ flow $J(x,t)$ from/to
$x=\pm \infty$ is negligible.
% FM  will be constructed here as the
% minimal theory in a sense that at every step we shall choose its
% variant with minimal number of DFs and theory constants.
% We'll consider here only the pure fuzzy states which aren't
% the probabilistic mixture of several different  states.

FM will be constructed here as  minimal theory in a sense that at
every step we shall choose the option with minimal number of DFs
 and theory constants. In such framework one should  look for
 $|g\}$ ansatz, which evolution would be linear.  Yet $m$ state representation in form
of the line: $|g\}=\{w, v\}$ is asymmetric relative to its norm
$w(x)$, and the evolution of its component $w$
 described by eq. (\ref {AZ2}) is nonlinear. Hence it's instructive to
look for symmetric $|g\}$ representation $\eta(x)$ for
 which its evolution is linear;  such  ansatz can be
  complex, quaternionic or some other symmetric form.
% It means that the expectation values of all $m$ observables
% $\{\bar {Q}\}$ should be some $g^0(x)$ functionals.
 Generally, in such framework  $\eta(x)=\Upsilon_x(w,v)$ where $\Upsilon_x$ is some
$w, v$ functional and $x$ is its parameter. However, $\eta$ norm
corresponds to $w(x)$, hence if $w(x) \to 0$ for some $x \to x_0$
, then $\eta(x)$ supposedly also should be negligible in $x_0$
vicinity. Hence $\eta$ can be decomposed also as:
\begin {equation}
    \eta(x)=\Upsilon_x(w,v)= f_r[w(x)] F_x(w, v)
                                     \label {AZ77}
 \end {equation}
 where $f_r$ is some real function, such that $f_r(\epsilon) \to 0 $
 for $\epsilon \to 0$; $F_x$ is an arbitrary functional.
%which, in principle, can be real, complex, quaternionic, etc..
In this vein, for $|g\}$  characterized by two DFs $w, v$,  it's
instructive to start $\eta$ ansatz search from complex $F_x$.

Plainly, $m$ evolution as the whole can be characterized by $m$
velocity $u(t)$ with expectation value $\bar {u}(t)$. Yet in FM,
alike $m$ coordinate $x$, such $u$ also can be also considered as
fuzzy value $\tilde {u}(t)$ with  corresponding normalized
distribution $w_{u}(u,t)$ which can be defined in $u$
measurements. Really, $m$ velocity measurement is related  to $x$
measurements in the different time moments $t,t'$, yet in FM the
fuzzy $\tilde x$ can possess some dispersion $\sigma_x(t)$ which
should result in appearance of corresponding $u$ uncertainty.
 Obviously, $\bar {u}(t)$ coincides
with
 $\bar{v}(t)$ of (\ref{AB}), hence:
\begin {equation}
   \bar{u}=\int\limits^{\infty}_{-\infty} v(x) w(x) dx
                                     \label {AY5}
 \end {equation}
 i.e. its value is also defined by $w, v$.
 In place of  $u$, below it will be convenient to use  the variable
  $p=\mu u$ where
  $\mu$ is the theory constant; for the corresponding distribution $w_p(p)$,
 % $$
 % \bar{u}=\frac{1}{\mu}\int pw_pdp=\frac{\bar{p}}{\mu}
% $$
it gives   $ w_u(u)=\mu w_p(\mu u)$.
% Meanwhile, $w_p(p)$ should be alsoLet's expressed as function of  $|g\}$ DFs
% $w(x),v(x)$,
% because  they should give the complete description of $m$ state $|g\}$.
 % Since $\eta(x)$ depends only on two DFs $w,v$, then
 % As was argued above, in the minimal case it's reasonable to look
% for $w_p$ expression via the complex functional.
If $|g\}$ is physical state then analogously to QM,  the
expectation value of arbitrary $m$ observable $Q$ in FM is
supposedly expressed as some $\eta$ functional.
%As was argued above, $w_p$ is supposedly defined  by $\eta(x),
 In particular,  $w_p(p)=F_p(\eta)$ where $F_p$ is
parameter dependent functional. The transformation $\eta \to w_p$
should possess the
 following properties:

 i) Norm conservation: if $\eta(x)$ is normalized, then the same is true for
 $w_p$.

 ii) $p$ expectation value $\bar p$ is expressed via $w(x),v(x)$
 according to eq. (\ref{AY5}).

 iii) For free $m$ evolution  $w_p$ is independent of $\eta(x) \to \eta(x+x_0)$ space
 shifts.\\
For complex  $\eta(x)$ it can be shown that its fourier transform
satisfies to these conditions \cite {Vlad}. To prove it and
calculate $w_p$,
  let's introduce  the auxiliary form:\\
  $ \varphi(p)=w_{p}^{\frac{1}{2}}e^{i\beta(p)}$, here $\beta(p)$
is the dummy  real function on which the final $w_p$ ansatz
 wouldn't depend. In accordance with  equality (\ref{AZ77}), $\eta(x)$ can be written
 as:
 \begin {equation}
   \eta(x)=f_r(w) G_x(w,v) e^{i\lambda_x(w,v)}
                                              \label {AY15}
 \end {equation}
 where
%  $f$ is the real function of $w$ in a given $x$;
$G_x(w,v),\,\lambda_x(w,v)$ are  real functionals.
  Consider then  $\varphi$
 fourier decomposition on $X$:
\begin {equation}
\varphi(p)  = \int\limits^{\infty}_{-\infty}\eta(x)e^{-ipx} dx=
\int\limits^{\infty}_{-\infty}
 f_r G_x e^{i\lambda_x-ipx} dx
                                                          \label{AQ}
\end {equation}
$w_p$ is normalized distribution, so the application of
Plancherele formulae to that norm gives \cite {Vlad}:
 \begin {equation}
\int\limits^{\infty}_{-\infty}w_{p}(p)dp  =
\int\limits^{\infty}_{-\infty} \varphi(p) \varphi^*(p)dp=
\int\limits^{\infty}_{-\infty} f_r^2(w)G^2_xdx =1 \label {AY25}
 \end {equation}
The calculation of $ \delta w$ variation for the equality:
\begin {equation}
   \int\limits^{\infty}_{-\infty} [f_r^2(w)G_x^2-w]dx =0
                                                         \label {AY35}
 \end {equation}
 demonstrates that $G_x$ doesn't depend on $v$. Thus it
permits to settle $G_x=1$ and $f_r(w)=\pm w^{\frac{1}{2}}(x)$.
Then $\bar p$ can be  calculated anew  from
 $2$-nd Plancherele formulae :
\begin {equation}
\bar{p}=\int\limits^{\infty}_{-\infty} p \varphi(p)
\varphi^*(p)dp= \int\limits^{\infty}_{-\infty} \frac{\partial
\lambda_x}{\partial x} f_r^2(w) dx
                   \label {AY55}
\end {equation}
% \cite{Vlad} and can be expressed as $w'(u)=|\varphi(u)|^2$ where
From the comparison with eq. (\ref {AY5})
% and $\bar{p}=\mu \bar u$
it follows: $\lambda_x= \gamma(x) + \chi(w)$ where $\gamma$ is the
functional:
 \begin {equation}
       \gamma(x)= \mu\int\limits^{x}_{-\infty} v(\xi) d\xi +c_{\gamma}  \label {AY}
 \end {equation}
 here  $c_{\gamma}$ is an arbitrary real number.
 $\chi(w)$ is an arbitrary real function which obeys to the
condition:
\begin {equation}
\int\limits^{\infty}_{-\infty}w \frac{\partial \chi}{\partial x}
dx=0
            \label {AY75}
 \end {equation}
and so is the analogue of $\eta$ gauge.
%One can choose for $f(x)$ an arbitrary sign,
% If to choose $f=w^{\frac{1}{2}}$,  then
As the result,  $w_{p}$ and $\beta(p)$
 can be calculated  from eq. (\ref {AQ}) as functions of
 $\chi$.
In particular:
\begin {equation}
       w_p(p)= |\int\limits^{\infty}_{-\infty} w^{\frac{1}{2}} e^{i\gamma+i\chi-ipx} dx|^2  \label {AYA}
 \end {equation}
is independent of $\beta({p})$,  so $w_p$ is really $w,v$
functional,
  on the all appearances for
minimal theory such $w_p$ ansatz is unique. $\beta(p)$ is, in
fact,  the analogue of $\gamma(x)$ for $p$ observable.
 %Below we shall use also the observable $p=\mu u$,
% if one substitute in the integral of eq. (\ref{AQ}) $ux$ by $px$ and $fe^{i\lambda_x}$
% Summing up these calculations, it supposes that the temptative
The resulting  $m$ state in $x$-representation:
%  is the fourier transform of $\varphi(p)$:
\begin {equation}
   \eta (x)=w^{\frac{1}{2}}(x)e^{i\gamma+i\chi} \label {AYA2}
\end {equation}
 is  the vector (ray) of complex Hilbert space $\cal H$. In this
framework, the observable $p$ corresponds to the operator
$\hat{p}=-i \frac{\partial }{\partial x}$ acting on $\eta$, i.e.
 $\bar{p}=\int \eta^*\hat{p} \eta dx$. Thus, $x$ and $p$
 observables are described by the linear self-adjoint operators,
which obey to the commutation relation $[\hat{x},\hat{p}]=i$.
% For any pair of linear, self-adjoint
% operators $\hat{Q}_1,\hat{Q}_2$ on $\cal H$ there is the unitary
% operator $\hat U_{2,1}$ such that $\hat {Q}_2= \hat{U_{2,1}}^{-1}
%\hat{Q}_1 \hat U_{2,1}$ \cite {Jauch}.
 By the analogy, we suppose that all $m$ PV
observables $\{ Q\}$ are the linear, self-adjoint operators on
$\cal H$.
% Such condition guarantees that that the observable
%expectation values are real.
If this is the case, $\eta(x)$ is the plausible candidate for
$|g\}$ state ansatz in $X$-representation, because for such $\eta$
the expectation values of all observables $\bar{Q}(t)$ should be
expressed as semi-linear $\eta$ functionals.

 Note that
$\eta=e^{i\chi}g$,
  where $g(x,t)$ is standard QM wave function, so that $\eta (x,t)$ is
  its trivial map. Thus we can study first
   $g(x,t)$ evolution, and then $\eta (x,t)$  properties
 will be derived basing on obtained results.
  Evolution equation for $g$  is supposed to be of the first order in time,
 i.e.:
  \begin {equation}
                  i\frac{\partial g}{\partial t}= \hat{H} g.      \label {AZ4}
 \end {equation}
 In general $\hat H$ can be nonlinear operator, for the simplicity
 we shall consider first the linear case and turn to nonlinear one
in the next section. Free $m$ evolution is invariant relative to
$x$ space shifts to arbitrary $x_0$ performed by the operator
  $\hat{W}(x_0)=\exp({x_0\frac{\partial}{ \partial x}})$.
 Because of it, the corresponding operator $\hat{H}_0$ should commute with
$\hat{W}(x_0)$ for the arbitrary $x_0$, i.e.
$[\hat{H}_0,\frac{\partial}{ \partial x} ]=0$. It holds only if
$\hat {H}_0$ is differential polinom
 % From the simple
%reasons its minimal rate $l \ge 2$,
 of the form:
 \begin {equation}
   \hat {H}_0=
%   \hat{H}_2 +  \hat{H}_n=
%  c_0 -c_1 \frac{\partial}{\partial x
    - \sum\limits_{l=1}^n b_l \frac{\partial^l }{\partial x^l}       \label {AU}
 \end {equation}
 where  $ b_l$ are  arbitrary real constants, $n \ge 2$.
From $X-$reflection invariance $b_l=0$ for noneven $l$. Suppose
that the action of external field on $m$ can be accounted in
$\hat H$ additively: $\hat {H}= \hat {H}_0+V(x,t)$ where $V$ is
real nonsingular function. Let's rewrite eq. (\ref {AZ4})
separating $w,\gamma$ derivatives:
\begin {equation}
                  i\frac{\partial g}{\partial t}=
(i\frac{\partial w^{\frac{1}{2}}}{\partial t}
  - w^{\frac{1}{2}} \frac{\partial \gamma}{\partial t})e^{i\gamma}=
   e^{i\gamma}\hat{Z}g      \label {AV}
 \end {equation}
% Let's compare the  left and right parts
%of eq. (\ref {AZ4}), for that purpose  $\hat H$ can be decomposed
%into $\hat{H}_2=-c_2\frac{\partial^2}{\partial x^2}$ and $\hat
%{H}_n=\hat{H}-\hat{H}_2$.
% }{\partial x^2}
% As follows from eq. (\ref {AZ4}), the operator $\hat{F}(g)$ is also equal to :
where $\hat{Z} = e^{-i\gamma}\hat{H}$. Hence:
\begin {equation}
\frac{\partial w^{\frac{1}{2}}}{\partial t}= im (\hat{Z}g) \label
                                                                 {AY85}
 \end {equation}
Yet if to substitute $v(x)$ by $\gamma(x)$ in eq. (\ref {AZ2}) and
transform it to ${w}^{\frac{1}{2}}$ time derivative, then:
\begin {equation}
\frac{\partial w^{\frac{1}{2}}}{\partial t}=
 -\frac{1}{\mu}  \frac{\partial
w^{\frac{1}{2}}}{\partial x}\frac{\partial \gamma}{\partial x}
 - \frac{1}{2\mu} w^{\frac{1}{2}} \frac{\partial^2 \gamma}{\partial
 x^2}     \label {BV}
\end {equation}
 Plainly, this expression and $im (\hat{Z}g)$ should coincide,
then $\hat H$  can be obtained from their comparison term by term.
% coincide up to $c_2/{\mu}$ ratio, from that $F_n$  can be derived.
  In particular, the imaginary part of $\hat{Z}g$  includes the highest
  $\gamma$ derivative as the
 term  $b_n w^{\frac{1}{2}} \frac{\partial^n \gamma }{\partial x^n}$, yet
 for  eq. (\ref{BV})
 the highest $\gamma$ derivative is proportional to
  $w^{\frac{1}{2}}\frac{\partial^2 \gamma }{\partial x^2}$.
  Hence it gives: $b_2=\frac{1}{2\mu}$ and for all $l>2$ it follows that $b_{l}=0$, only in this case both
   expressions  for $\frac{\partial w^{\frac{1}{2}}}{\partial t} $ would coincide.
%   the same is true for other $\hat{Z}g$ terms.
  Thus $g$ free evolution is described by the only $\hat {H}_0$
  term  $b_2=\frac{1}{2\mu}$, so $\hat H$ is  Schroedinger hamiltonian
  for  particle with mass $\mu$.
% The temptative interpretation of this
% result is that the flow continuity equation of (\ref {AZ2}) is
% incompatible with $g(x)$ dynamics which would depend  on $g$
% derivatives with the rate higher than $2$.
% with arbitrary $c_0$ which can be taken to be zero.
     The obtained ansatz
gives also $J(\pm\infty,t)=0$ for $w$ flow of eq. (\ref{AZZZ}), in
accordance with our assumptions.  Note that in standard QM $m$
evolution equation is, in fact, postulated $ad\,hoc$; here it's
derived from FT premises for particle evolution on fuzzy
manifold. The same is true for the commutation relation
$[\hat{x},\hat{p}]=i$

In this framework  the flow velocity $v(x,t)$ isn't observable,
but can be formally defined as the ratio of $J(x),w(x)$
observable expectation values, where $w$ observable is described
by the projection operator $\hat {\Pi}(x)$; the operator
$\hat{J}(x)$ considered in \cite {Schiff}. As was noticed earlier,
the particle evolution in QM in some aspects is similar
 to the  motion of continuous media \cite {Lan}.
 This analogy is explored thoroughly in
hydrodynamical QM model (QFD) \cite {Made,Gh},
 its connection with FM will be discussed in sect. 5.

Plainly, $\gamma(x)$ corresponds to $|g\}$ quantum phase,
  so that:
   $$
   k(x,x')=\gamma(x)-\gamma(x')
   $$
    describes the  phase
   correlation between the state components in $x, x'$.
%   Such correlations induce, in fact, the interference effects between $|g\}$
% components in $x,x'$. As was noticed above,
Thus pure FM state can be characterized by the density $w(x)$  and
the array of  bilocal geometric correlations $\{\kappa_l(x,x')\}$,
the first of them: $\kappa_0 (x,x')$ was introduced in sect. $2$.
% so that $K_1(x,x')$ can be also regarded as  $k$ component.
 Until now
we've considered only the pure fuzzy states, i. e. the states
which aren't the probabilistic mixture of several pure
 states.
Analogously to QM, the  mixed states in FM can be defined via the
density matrixes, i.e.  the positive, trace one operators $\rho$
on $\cal H$ \cite {Jauch}.
% The
% purity rate of mixed states  in $x$-representation $\rho(x,x')$
% corresponds to  $K_0(x,x')$.
 In particular, for pure $m$ states:
$$
   \rho(x,x')=g(x)g^*(x')=[w(x)w(x')]^{\frac{1}{2}}e^{ik(x,x')}
$$
is equivalent to $g(x)$, yet
 such $|g\}$ representation demonstrates in the open
 the  correlation structure of $m$ pure states.
% As was noticed in sect. 3, in FM
Thus, the most consistent FM state  ansatz is given by the density
matrix $\rho$. However,
%the direct derivation of $\rho$
the evolution equations for pure states in form of $\rho$  are
more complicated then for Dirac vector $g(x)$, and because of it,
we shall exploit it throughout our paper.

% Such correlations induce, in fact, the interference effects
% between $|g\}$ components in $x,x'$. The mixed states in FM are
% defined exactly like in QM formalism, i.e. are positive, trace
% one operators $\rho$ on  $\cal H$. The purity rate of mixed
% states  in $x$-representation $\rho(x,x')$ corresponds to the
% weight correlation $K_f(x,x')$ defined in sect. 2. For pure $m$
% states:
% $$
%   \rho(x,x')=g(x)g^*(x')=[w(x)w(x')]^{\frac{1}{2}}e^{i\Delta(x,x')}
% $$
% is equivalent to $g(x)$, yet
% such $|g\}$ representation demonstrates in the open
%  the  correlation structure of $m$ pure states.

% \section {General Fuzzy Dynamics}

\section{ General Fuzzy Dynamics }
\label{sec-2}

 In the previous section  $1$-dimensional FM formalism
  was derived from FT premises
 assuming that $|g\}$ evolution is linear
 and $|g\}$ gauge $\chi(w)$ can be neglected. Now
  these assumptions will be dropped one by one and the general formalism
   derived.
 Concerning with nonlinear evolution, the conditions of QM
 linearity were reconsidered by Jordan, and turn out to be essentially
 weaker than Wigner theorem asserts  \cite {Jor}.
  In particular, it was proved that if the evolution
 maps the set of  all pure states one to one onto itself, and for
 arbitrary
  mixture of orthogonal states $\rho(t)=\sum P_i(t)\rho_i(t)$ all
 $P_i$ are independent of time,
  then  such evolution is
  linear. Here $\rho_i(x,x',t)=g_i(x,t)g^*_i(x',t)$ are the density matrixes
 of orthogonal pure states $g_i$.
 Yet for the considered FM formalism  first condition is,
  in fact, generic: no mixed  (i.e. probabilistic) state
 can appear in the  evolution of pure fuzzy state. The second condition
 involves the probabilistic mixture of such orthogonal states
  and also seems to be rather  weak assumption.
 % and so FM evolution should be linear.
%The transition to 3-dimensional case is trivial.
%Overall, the free evolution of $m$ fuzzy state $|g(t)\}$
% is described by Schroedinger equation with free
% Hamiltonian $\hat{H_0}=\frac{\hat{p}^2}{2m_0}$.

Now let's return to  $\eta(x)$ ansatz of (\ref {AYA2}), it can be
shown  that Jordan theorem demands also  that $\chi(w)=0$.
 For  $m$ states of (\ref {AYA2}) and corresponding $g$ ansatz,
  if $\langle g_i|g_j\rangle=\delta_{ij}$, then
$\langle \eta_i|\eta_j\rangle=\delta_{ij}$ and vice versa. As was
argued above, in FM any pure state $g(t_0)$ should evolve to pure
state $g(t)$ for arbitrary $t$, so the same should be true for
any $\eta(t_0)$.
% Thereon, $g'$ evolution equation supposedly can be
% also expressed as: $i\frac{\partial g'}{\partial t}= \hat{H'} g'$,
% yet here $\hat H'$ isn't necessarily linear.
  Now  Jordan theorem can be
 applied  to $\eta$ evolution, to demonstrate it  consider $g$ evolution
 equation:
%  Let's decompose $i\frac{\partial }{\partial t}$
% analogously to eq. (\ref{AV}):
\begin {equation}
i\frac{\partial g}{\partial t}=
 i\frac{\partial }{\partial t}(\eta e^{-i\chi})=
i\frac{\partial \eta}{\partial t}e^{-i\chi}+ \eta \frac{\partial
\chi}{\partial w} \frac{\partial w}{\partial t}e^{-i\chi}=
 \hat{H}(\eta e^{-i\chi})       \label {AVV}
\end {equation}
From  it one can come to the equation for $\eta$, the
term containing $\frac{\partial w}{\partial t}$ can be rewritten
according to (\ref{BV}). As the result,
it gives:
\begin {equation}
                  i\frac{\partial \eta}{\partial t}=
e^{i\chi}\hat{H}(\eta e^{-i\chi})+
\frac{\eta}{\mu}e^{i\chi}\frac{\partial \chi}{\partial w}
\frac{\partial }{\partial x}(w \frac{\partial \gamma}{\partial
x})     \label {AV3}
\end {equation}
Resulting equation for $\eta$ is also of first time order, but is
openly nonlinear. Therefore, for arbitrary $\chi(w)$, given the
initial $\eta(x,t_0)$, the resulting $\eta (x,t)$ is just the
equivalence class of $g(x,t)$ which evolves linearly from
%\begin {equation}
$g(t_0)=\eta(t_0)e^{-i\chi}$.
% \label {AV39}
%\end {equation}

% In FM $x$ is $m$ observable and it's sensible to admit that
%$\hat p_x=i\frac{\partial}{\partial x}$ describes $m$ momentum and
%all self-adjoint operator functions $\hat {F}_Q(x,p)$
%  So in FM all physical observables constitute
% the linear algebra of self-adjoint operators (Segal algebra).
%
%In our approach the partial ordering of material point $m$
%induces the nonlocalized density distribution $w(x)$, its
%evolution is similar to the motion of compressed liquid \cite
%{Lan}. In particular, second equation of system (\ref {AZ22}) is
%similar to Eiler equation where first term corresponds to its
%internal pressure or diffusion.

 %   \section {3-dimensional FM Dynamics}

%   Here we shall consider

 FM  for $3$-dimensional geometry, in fact,  doesn't demand any
 principal modification of described formalism. In $3$ dimensions our
 fundamental set $C^F=A^p \cup R^3$,
 % hence for any fuzzy point $a_j \in A^p$ its
 % ordering  properties should be  defined relative to $X,Y,Z$
 % coordinate axes
 % separately. Assuming FM rotational invariance, it follows that
 hence FP $a_j$ fuzzy properties are described by the positive function
  $w^j(\vec{r})$ with norm $\int w^j d^3r=1$.
 If  the particle $m$ corresponds to the fuzzy point $a(t)$ characterized by
 $w(\vec{r},t)$, then
 analogously to sect. $2$, given  $w$ evolution depends
  on local parameters only,  it can be expressed as:
 \begin {equation}
                  \frac{\partial w}{\partial t}(\vec{r},t)=-\Phi(\vec{r},t)     \label {AXY3}
 \end {equation}
where $\Phi$ is an arbitrary local function.  From $w$ norm
conservation it follows that:
 \begin {equation}
   \int\limits_{W} \Phi(\vec{r},t)d^3r =
   \int\limits_{W} \frac{\partial w}{\partial t}(\vec{r},t)d^3r=
       \frac{\partial }{\partial t}\int\limits_{W}
       w(\vec{r},t)d^3r = 0
                         \label {AZZ3}
 \end {equation}
 where $W$ denotes the infinite volume with $|\vec{r}| \to \infty$ in all directions.
 Analogously to sect. $2$, we suppose that  $\Phi$ integral
counterpart $\vec J$ exists and defined via the relation:
$f=div\vec{J}$ where $\vec J$ is some vector function. Then  eq.
(\ref{AZZ3}) is fulfilled if:
 \begin {equation}
   \int\limits_{S} \vec{J}\vec{n}ds=0     \label {Y}
\end {equation}
where $S$ is the surface surrounding $W$, $\vec n$ is the  vector
normal to the given surface element. If this is the case, then
eq. (\ref {AXY3}) can be transformed to flow continuity equation:
\begin {equation}
                  \frac{\partial w}{\partial t}=-div\vec{J}     \label {AXY5}
 \end {equation}
One can decompose $\vec{J}=w\vec{v}$ and consider $w$ flow
velocity $\vec{v}(\vec{r})$ as independent $|g\}$ parameter. $m$
state $|g\}$ is supposed to be the complex $w, \vec v$ functional
$g(\vec{r})=\Upsilon_{\vec{r}}(w,\vec{v})$. For $m$ as the whole,
its velocity is supposedly characterized by  fuzzy vector
 $\tilde {\vec u}$ which corresponds to distribution  $w_u(\vec
 {u})$, so that:
\begin {equation}
\langle \vec{u} \rangle= \int \vec{u} w_u(\vec{u}) d^3u= \int
\vec{v}(\vec{r})w(\vec{r}) d^3 r                    \label {AX25}
 \end {equation}
$m$ kinematical  fuzzy momentum defined as: $\vec{p}=\mu \vec{u}$.
 From that, analogously to eqs.
(\ref{AY5} - \ref{AYA2}),  standard QM ansatz for $m$ state
obtained: $g(\vec{r})=w^{\frac{1}{2}}e^{i\gamma}$ where
 $g$ phase $\gamma(\vec{r})$ obeys to
the equality $\mu\vec{v}=grad(\gamma)$. To guarantee  the formalism
consistency, we assume that  the phase
correlation value
 $K_1(\vec{r},\vec{r'})$ is
 independent of the path $l$ between  $\vec r$, $\vec r'$ over which it can be
 calculated additively :
\begin {equation}
K_1(\vec{r}, \vec{r'})=\gamma(\vec{r})-\gamma(\vec{r}')=
   \int\limits_{\vec{r'}}^{\vec{r}} grad(\gamma) d\vec{l} \label {U666}
\end {equation}
%
% has the same value for arbitrary path $l$.
% In other terms it means that for arbitrary
% $\vec r$ the vorticity $W(\vec{r})=0$, where $W=curl (grad\gamma)$ \cite {Gh}.
%Note that in
%hydrodynamics this condition corresponds to the fluid potential
%motion \cite {Lan}.
% aaa $\gamma(\vec{r})-\gamma(\vec{r}')=$

% From the similar sequence  of calculations, as for $1$-dimensional case,
%Here we shall exploit the alternative method of  free
%%free Schroedinger equation can be  derived for $3-$dimensional
%geometry.
 Considering $g$  evolution,
% the start  that
%  $g$ evolution operator $\hat H$ is linear. Then
 for free $m$ linear
evolution its operator $\hat{H}_0$ should be the even polinom of the form:
\begin {equation}
   \hat {H}_0=
%   \hat{H}_2 +  \hat{H}_n=
%  c_0 -c_1 \frac{\partial}{\partial x
    - \sum\limits_{l=1}^n b_{2l} \frac{\partial^{2l} }{\partial \vec{r}^{2l}}
                                                              \label {AU9}
 \end {equation}
If the external field action can be described by the addition of real function $V$ to it:
\begin {equation}
\hat {H}=\hat{H}_0+V(\vec{r},t)
                                    \label {UU99}
 \end {equation}
 then
from $\frac{\partial g}{\partial t}$ the term $\frac{\partial
w^{\frac{1}{2}}}{\partial t}$ can be extracted and expressed via
$w,\gamma$ $\vec r$-derivatives. From their comparison with
corresponding $\hat Hg$ derivatives
% it follows
%
% $b_2=\frac{1}{2\mu}$ and $b_{2l}=0$ for $l>1$, i.e.
  Schroedinger equation is obtained for $m$ evolution.
% Note that for all normalized states the condition (\ref {Y}) is
%   fulfilled, because for them $\vec{J}(\vec{r}) \to 0$
%    at $\vec{r} \to \infty$.
The applicability of Jordan theorem to $3$-dimensional $\hat H$
is obvious, because the derivation of $\hat H$ linearity doesn't depend on
 the dimensiality of coordinate space. The same is true for the proof of
uniqueness of $g(\vec{r},t)$ ansatz, i.e. that $\chi(w)=0$.
% which can be performed analogously to
%  the  calculation sequence given by  eqs. (\ref{AVV8}-\ref{AV3}).
% 1-domensional considerations of sect. 2.
%All  $m$  states $g(\vec{r},t)$ belong to $\cal H$, hence the
%superposition principle holds true in FM also.
%  We shall admit that in FM the observer can prepare an arbitrary
%  initial state $g_0(x)$, such assumption corresponds to the
%  superposition principle, because $g^0$ can be decomposed
%  $g_0=g^0_1+g^0_2$ where $g_{1,2}$ are arbitrary states with
%  appropriate weights.

 In our derivation of evolution
equation we didn't assume Galilean invariance of FM, rather in
our approach it follows from the obtained evolution equation if
the reference frame (RF) is regarded as the physical object with
mass
 $\mu \to \infty$ \cite {May2}.
 For the transition to  relativistic FM from our ansatz
 its natural extension for complex scalar state $g$ is
  Klein-Gordon equation.
% evolution becomes the important criteria for the choice of
% consistent ansatz. If to
Yet for such equation it's impossible to define $m$ probability
density $w(\vec{r})$  which would be
  nonnegative for all free states \cite {Jauch}.
As was noticed in sect. $3$, in principle, $m$ scalar state can be
complex, quaternionic, octonionic, etc..
% Hence to cope with this
%difficulty, one can try  to replace in Klein-Gordon equation the
%complex $g\vec{r})$ by some other scalar.
 We find that the
minimal consistent $|g\}$ ansatz gives  quaternion scalar
$\xi(\vec{r})$, so that:
\begin {equation}
(\frac{\partial^2}{\partial t^2}-\bigtriangleup +\mu^2)\xi=0
 \label {AU29}
 \end {equation}
For such state the single quantum $g$ phase $\gamma$ is extended
to three  independent phases $i\gamma+j\beta+k\alpha$ which
correspond to additional geometric DFs.
%  Such DFs can be
% considered as $m$ phase correlations $K_{1,2,3}$ analogous to
 (\ref {U666}).
 To get nonnegative
$w (\vec{r})$ one should broke first $i,j,k$ space symmetry and to
choose an arbitrary preferred  basis $i',j',k'$.  Plainly, in
this basis $\xi=\psi_1+\psi_2 j'$, here
$\psi_{1,2}=a_{1,2}+b_{1,2}i'$ where  $a_{1,2}, b_{1,2}$ are real
functions. Let's rewrite $\psi_{1,2}$ in form of spinor $\varpi_u$
and define the auxiliary  spinor $\varpi_d$:
\begin {equation}
\varpi_d=-i'(I\frac{\partial }{\partial t}+ {\frac{\partial
}{\partial \vec{r}}} \vec{\sigma})\varpi_u
                                                        \label {AU299}
 \end {equation}
in obvious notations.
% If to denote $\psi_d$ components as $\psi_{3,4}$,
If to denote up, down $\varpi_d$ components as $\psi_{3,4}$, then
 it's easy to check that $w(\vec{r})=\sum_1^4|\psi_l|^2$
is nonnegative and normalizable for  arbitrary $\xi$. If to regard
$\psi_{1,...,4}$ as $4$-spinor components, then such $4$-spinor
obeys to Dirac equation in chiral representation \cite {Fay}.
Hence such $w(\vec{r})$  would evolve according to  flow
continuity equation stipulated by Dirac equation. It seems that in
FM some geometric DFs can be 'compactified', resulting in the
appearance of internal spinor space, so that the particle $m$
acquires spin $\frac{1}{2}$.

Now we shall consider   the interaction between fuzzy states in
FM framework.
% and discuss their possible generalization for
%    the relativistic case.
 Note first that in FM by derivation  the free Hamiltonian $\hat{H}_0$
induces, in fact, $\cal H$ dynamical asymmetry between
 $|\vec{r}\rangle$ and $|\vec{p}\rangle$ 'axes'.
% which $a\, priory$ is absent in standard QM formalism.
% By itself,  FM description of physical systems doesn't need
% the corresponding classical system as the starting point \cite {Bl}.
%% Copenhagen QM interpretation claims that
%% QM can't be formulated consistently without the preliminarily postulated
%% classical notions, it seems that FM formalism is, at least,
%% essentially less connected with them.
% Hence in our approach only
% some general properties of  classical dynamics will be implemented
% but no particular correspondence principle is used.
 As follows from eq.( \ref{AU}-\ref{BV})  $m$ free dynamics can be
described by the system of two equations which define
$\frac{\partial {w}^{\frac{1}{2}}}{\partial t}$ and
$\frac{\partial\gamma}{\partial t}$ which for $3$-dimensions are
equal to:
$$
\frac{\partial w^{\frac{1}{2}}}{\partial t}=
 -\frac{1}{\mu}  \frac{\partial w^{\frac{1}{2}}}{\partial \vec{r}}\frac{\partial \gamma}{\partial \vec{r}}
 - \frac{1}{2\mu} w^{\frac{1}{2}} \frac{\partial^2 \gamma}{\partial {\vec r}^2}
$$
 \begin {equation}
\frac{\partial\gamma}{\partial t}= -\frac{1}{2\mu}
[(\frac{\partial\gamma}{\partial \vec{r}})^2-
  \frac{1}{w^{\frac{1}{2}}} \frac{\partial^2 w^{\frac{1}{2}}}{\partial \vec {r}^2}]  \label {AZ22}
\end {equation}
%The similar equations are exploited in QFD formalism \cite{Gh}.
Yet  the first of them is equivalent to eq. (\ref{AXY5}) which
describes just $w(\vec{r})$ balance and so is, in fact,
kinematical one and can't depend on any  interactions directly.
Namely, under some external influence the values of $w, \gamma$
variables can change, but no new terms can appear in that
equation. Note that in QM $e\vec A$ term formally appears in it,
but it's just the part of the  expression for kinematic momentum
\cite {Jauch}.
% Namely, by the analogy with QM, in FM one can admit:
% \begin {equation}
% \frac{\partial\gamma}{\partial \vec{r}}= \mu  \vec{v}-q_1\vec{A}
%  \frac{1}{w^{\frac{1}{2}}} \frac{\partial^2 w^{\frac{1}{2}}}{\partial \vec {r}^2}]
%    \label {AZ78}
% \end {equation}
% where $q_1$ is $m$ electric charge.
 Hence $m$ interactions can be accounted only via the
modification of  second equation of system (\ref{AZ22}).
 Assuming that analogously to eq. (\ref {UU99}) the evolution terms are real and additive, it gives:
$\hat{H}=\hat {H}_0  +\hat{H}_{int}$
% it gives:
% \begin {equation}
%\frac{\partial\gamma}{\partial t}= -\frac{1}{2\mu}
%[(\frac{\partial\gamma}{\partial \vec{r}})^2-
%  \frac{1}{w^{\frac{1}{2}}} \frac{\partial^2 w^{\frac{1}{2}}}{\partial \vec %{r}^2}]
%   +\hat{H}_{int}  \label {AZ99}
%\end {equation}
where $\hat{H}_{int}$ is the interaction term.
% which is nontrivial
%if $\frac{\partial\hat{H}_{int}}{\partial \vec{r}} \ne 0$.
Let's consider how the interaction of two particles $m, M$ can be
described in such approach. Suppose also that $m, M$ interaction
is universal in a sense that
 $\langle\hat{H}_{int}\rangle \ne 0$ for arbitrary relative $m,M$
 momentum $\langle \vec {p}_{12} \rangle$, and is induced by the conserved
 charges $q_1,q_2$. Then the main $\hat{H}_{int}$ term which survives
 at $\langle\vec {p}_{12}\rangle \to 0$ is equal to $q_1q_2U(r_{12})$.
as the result, $U(r_{12})$ corresponds to the classical potential.
In standard QM such interaction is, in fact, postulated from
classical-to-quantum correspondence, whereas here it follows from
FM geometric premises. Since $\gamma$ corresponds to the quantum
phase, it supposes that in FM  $m$ interactions can possess some
form of local gauge invariance \cite {Li}.
% Of course, one can
% just postulate the gauge interactions of certain kind, yet is
% seems worth to explore whether such dynamics can be obtained from
% some considerations related to FM or some other fundamental
% principles. In our previous paper the toy-model of abelian gauge
% interactions on fuzzy manifold was formulated which in the main
% aspects is similar to  QED \cite {Vax}.
% Despite that the fermion
% %state is described by several quantum phases, the same invariance
% fulfilled for it and can be extended also on relativistic case.
% Preliminary results for

% vector $|g\}$, so it was used throughout our paper.

\section {Conclusion}

It's well known that QM can be described by several alternative
formalisms, of them the most notorious are algebraic QM and
Schroedinger or standard formalism. To discuss the possible
advantages of FM formalism it's instructive to compare it with the
latter one.  From the formal side, standard QM exploits two
fundamental structures of different nature: space-time manifold
$R^3*T$ and functional space $\cal H$ defined on $R^3$. In
distinction, FM formalism involves only one basic structure, it's
fuzzy manifold $\tilde{R}^3*T$. FM physical states are
$\tilde{R}^3$ points, their equivalence to $\cal H$ Dirac vectors
was proved here.
% Such simplification makes the theory mathematically minimal
% and more integral.
In standard QM the evolution equation or postulated $ad\, hoc$ or
derived assuming Galilean invariance of object states \cite
{Jauch}. In FM the Schroedinger equation is derived assuming only
space-time shift invariance which is essentially weaker
assumption. Besides, the quantum-classical transition in such
theory is essentially more simple, it's just the transition of
$\tilde{R}^3$ manifold to $R^3$ one, for which the classical
particles correspond to ordered  points.  As the result, FM
formalism possesses simple and logical axiomatics which origin is
basically geometrical. It permits, in principle, to explore under
 new angles those quantum theories  for which geometry is the
formalism cornerstone, first of all these are quantum gravity and
gauge fields.

Concerning with the connection between FM and QFD noticed in sect.
2, in the latter theory all QM axioms are accepted at the initial
stage. Then, the postulated Schroedinger equation is rewritten as
the system of equations for the motion of classical liquid,
similar to eq. (\ref {AZ22}) \cite {Gh}. Yet at the next stage,
to reach the complete classicality of the theory some $ad\, hoc$
assumptions are added, however the resulting theories contradict
to experimental results. In comparison, FM is principally
nonclassical theory,  this nonclassicality originates from the
novel topological structure of space-time, whereas in standard QM
formalism the space-time geometry is the same, as in classical
mechanics.

% It's worth to notice also that  in such formalism the commutation
% relation $[\hat{x},\hat{p}_x]=i$  results, in fact, from the
% geometry and topology of fuzzy manifold.

 In our approach the state space is defined by geometry and
corresponding dynamics i.e. is derivable concept. For pure states
 of free nonrelativistic particle $m$
 it obtained to be equivalent to $\cal H$, but, in principle,  it can be different for other
systems. The similar features possess the formalism of algebraic
QM where the state space is defined by the observable algebra and
system dynamics \cite {Jauch}.
% FM observables coincide copiously with QM observables.
% As was noticed in sect. 3, in FM the most consistent state  ansatz
% is given by the density matrix $\rho$. However, the direct
% derivation of $\rho$ evolution equation is more complicated then
% for Dirac vector $|g\}$, and because of it, we used $|g\}$
% throughout our paper.
   Planck constant $\hbar=1$ in our FM
ansatz, but the same value ascribed to it in relativistic unit
system in which the velocity of light $c=1$; in FM framework
$\hbar$ only connects $x,p$ geometric scales and doesn't have any
other meaning.

In conclusion, we have shown that the quantization of elementary systems
 can be derived directly from axiomatic of set theory and topology together
with the natural assumptions about system evolution.
%% Namely, the commutative
%% fuzzy space $C^F$, in fact,  induces the $x,p_x$ noncommutativity
%% due to obvious uncertainty of $m$ velocity $\dot{x}(t)$ in such space.
 It allows to suppose that the quantization phenomenon has its roots
 in foundations of mathematics \cite {Jauch}. Our approach permits
 to construct QM formalism starting from geometric concepts and
 structures only, so in these aspects it's analogous to general
relativity construction.
In the same time the considered fuzzy manifold describes the
possible variant of fundamental pregeometry which is basic
component of some quantum gravity theories \cite {Bal}.
%  Note
% also that in such geometry the fundamental set of elements is, in
% fact, absent, it replaced by the set of positive functions, which
% makes it similar to the basic structure of noncommutative geometry
% \cite {Con,Bal}.
% The main aim of our theory, as well as  other studies of fuzzy
% spaces, is the construction of nonlocal QFT (or other more
% general theory) \cite {Bl}.
 In this vein,
FM provides the interesting opportunities, being generically
nonlocal theory which, in the same time, can possess Lorentz
covariance and  local gauge invariance.
%% We shall suppose that such interaction
%%  doesn't extinct even if the relative momentum of two particles
%% $|\vec {p}_{1,2}| \to 0 $ and  is related to some charge.
% g Yang-Mills fields \cite {Li}.

%\end {document}
% **********************************************************

\end {document}